\documentclass{article}
\usepackage{adjustbox}
\usepackage{spconf,amsmath,graphicx}
\usepackage{hyperref}
\usepackage{booktabs,multirow}
\usepackage{enumitem}
\usepackage{comment}

\title{Investigating salient representations and label Variance in Dimensional Speech Emotion Analysis}
\name{Vikramjit Mitra, Jingping Nie, Erdrin Azemi}
\address{Apple}
\begin{document}

\setlength{\abovedisplayskip}{3pt}
\setlength{\belowdisplayskip}{3pt}

%
\maketitle
\begin{abstract}
Representations derived from models such as BERT (Bidirectional Encoder Representations from Transformers) and HuBERT (Hidden units BERT), have helped to achieve state-of-the-art performance in dimensional speech emotion recognition. Despite their large dimensionality, and even though these representations are not tailored for emotion recognition tasks, they are frequently used to train large speech emotion models with high memory and computational costs. In this work, we show that there exist lower-dimensional subspaces within the these pre-trained representational spaces that offer a reduction in downstream model complexity without sacrificing performance on emotion estimation. In addition, we model label uncertainty in the form of grader opinion variance, and demonstrate that such information can improve the model’s generalization capacity and robustness. Finally, we compare the robustness of the emotion models against acoustic degradations and observed that the reduced dimensional representations were able to retain the performance similar to the full-dimensional representations without significant regression in dimensional emotion performance.
\end{abstract}
\section{Introduction}
\label{sec:intro}

Speech-based emotion estimation models aim to estimate the emotional state of a speaker from their speech utterances. Speech emotion models have garnered much attention to augment existing human-machine interaction systems, where such systems can not only recognize the words spoken by the speaker, but can also interpret how such words are expressed. Real-time speech-emotion models can help to improve human-computer interaction experiences, such as voice assistants \cite{mitra2019leveraging, kowtha2020detecting} and facilitate health/wellness applications such as health diagnoses \cite{desmet2013emotion, stasak2016investigation}.

Speech emotion research has pursued two distinct definitions of emotion: (1) \textit{discrete emotions}, and (2) \textit{dimensional emotions}. \textit{Discrete emotions} are categorized as, for example, fear, anger, joy, sadness, disgust, and surprise \cite{ekman1992argument}. These categories of emotions suffer from a lack of an agreed-upon lexicon of standard emotions, where the choice of emotions can vary \cite{ekman1992argument, cowen2017self}. Variation in discrete emotion definitions may result in annotation complexities, difficulty in realizing datasets with consistent emotion labels, and failure to include rare and/or complex emotional states. \textit{Dimensional emotion} \cite{posner2005circumplex} helps to systematically represent the emotion space using a 3-dimensional model of \textit{Valence}, \textit{Activation} (or, Arousal) and \textit{Dominance}, which maps discrete emotional states to a point in the 3-dimensional continuous space. In \textit{dimensional emotion}, \textit{Activation} reflects the energy in the voice, \textit{Valence} indicates negativity versus positivity, and \textit{Dominance} specifies how strong or submissive one may sound. Early studies on speech-based emotion detection focused on acted or elicited emotions \cite{busso2008iemocap}, however, models trained with acted emotions often fail to generalize for spontaneous subtle emotions \cite{douglas2005multimodal}. Recently, attention has been given to datasets with spontaneous emotions \cite{mariooryad2014building}, however, ground-truth labels for such data are often noisy due to varying degrees of grader agreement. Label uncertainty due to the variance in grader agreements is addressed by taking the mean of the grader decisions, however, the mean fails to account for examples that are challenging (where graders seem to mostly disagree with one another) compared to the samples that seem to be easier (where most graders agree). Modeling such uncertainty \cite{prabhu2022label} can be useful in speech emotion modeling to account for audio samples that were perceptually difficult to annotate. In this work, we investigate incorporating label variance into emotion modeling and demonstrate a gain in performance compared to the model not using label variance.

Combining lexical- and acoustic-based representations has been shown to boost model performance on emotion recognition from speech \cite{sahu2019multi, ghriss2022sentiment, srinivasan2021representation, siriwardhana2020jointly}. Recent studies have shown that using representations from pre-trained models can improve emotion recognition performance \cite{siriwardhana2020jointly, srinivasan2021representation, mitra2022Speechemotion, mitra2023pre}. However, such pre-trained representations are high dimensional, which poses both computational and memory challenges for the downstream emotion modeling task. It is also an open question, whether all the dimensions in such a large dimensional representation are needed for emotion analysis tasks or if there exists a low-dimensional subspace that contains the relevant affect-specific information. In this work, we investigate obtaining a low-dimensional subspace of pre-trained representations for emotion modeling and demonstrate that reducing the input size does not result in significant regression of emotion model performance, but helps to reduce the model size substantially. Note that the focus of this work is to reduce the downstream emotion model's size, not the pre-trained model sizes. In this work, we demonstrate:
\vspace{-1mm}
\begin{enumerate}[leftmargin=*]
\item Modeling label variance can help to achieve \textit{state-of-the-art} dimensional emotion estimation performance.
\vspace{-2mm}
\item Not all dimensions of the pre-trained model representation are needed for the emotion modeling task.
\vspace{-2mm}
\item Model size can be reduced by reducing input feature dimension, without regression in model performance. 
\end{enumerate}
\vspace{-4mm}

\section{Data}
\vspace{-2mm}
We have used the MSP-Podcast dataset (ver. 1.6) \cite{mariooryad2014building, lotfian2017building} that contains speech spoken by English speakers. The speech segments contain single-speaker utterances with a duration of 3 to 11 seconds. The data contain manually assigned valence, activation and dominance scores (7-point Likert scores) from multiple graders. The data split is shown in Table \ref{tab:table1}. To make our results comparable to literature \cite{ghriss2022sentiment, srinivasan2021representation}, we report results on Eval1.3 and Eval1.6 (see Table \ref{tab:table1}). To analyze the robustness of the emotion models, we add noise and reverberation to the Eval1.6 set at 3 SNR levels, resulting in three additional evaluation sets (see Table \ref{tab:table1}). 

\vspace{-4mm}
\begin{table}[h!]
\centering
\caption{MSP-podcast data split and noise-degraded sets.}
\vspace{1mm}
\begin{tabular}{lll}
\hline
\textbf{Split} & \textbf{Hours} & \textbf{Description}\\
\hline
Train1.6 & 85 & Training set \\
Valid1.6 & 15 & Validation set \\
Eval1.3 & 22 & Podcast1.3 evaluation set \\
Eval1.6 & 25 & Podcast1.6 evaluation set \\
$Eval1.6_{25dB}$ & 25 & Eval1.6 + noise within 20-30 dB \\
$Eval1.6_{15dB}$ & 25 & Eval1.6 + noise within 10-20 dB \\
$Eval1.6_{5dB}$ & 25 & Eval1.6 + noise within 0-10 dB \\
\hline
\end{tabular}
\label{tab:table1}
\end{table}
\vspace{-4mm}

\section{Representations \& Acoustic Features}
\vspace{-2mm}
\subsection{Acoustic Representations from HuBERT}
\vspace{-2mm}
We explore embeddings generated from HuBERT large, a pre-trained acoustic model \cite{hsu2021hubert}, which was pre-trained on 60K hours of speech from the Libri-light dataset with 24 transformer layers and 1,024 embedding dimensions. In our study, we extracted the following embeddings: 
\vspace{-2mm}
\begin{enumerate}[leftmargin=*]
\item $HUBERT_L$ embeddings from the $24^{th}$ layer of the pre-trained HuBERT large model (no fine-tuning).
\vspace{-2mm}
\item $HUBERT_{A}$ embeddings ($24^{th}$ layer) from $HUBERT_L$ model fine-tuned on 100 hours of Librispeech data. 
\end{enumerate}
\vspace{-3mm}

\subsection{Lexical Representations from BERT}
\vspace{-2mm}
We have used pre-trained Bidirectional Encoder Representations from Transformers (BERT) \cite{kenton2019bert} model to extract lexical embeddings for dimensional emotion model training. We rely on two pre-trained ASR models to estimate transcripts from speech: $HUBERT_{A}$ (mentioned above), and an in-house ASR system. Given transcribed speech, we normalize the text (removing punctuation \& considering all characters in lowercase) and use the resulting data as input to pre-trained BERT that generates 768-dimensional utterance-level embeddings from the $12^{th}$ layer. 

\subsection{Salient representations}
\label{sec:format}
\begin{figure}[h!]
\begin{minipage}[b]{1.0\linewidth}
  \centering
  \centerline{\includegraphics[width=7.5cm]{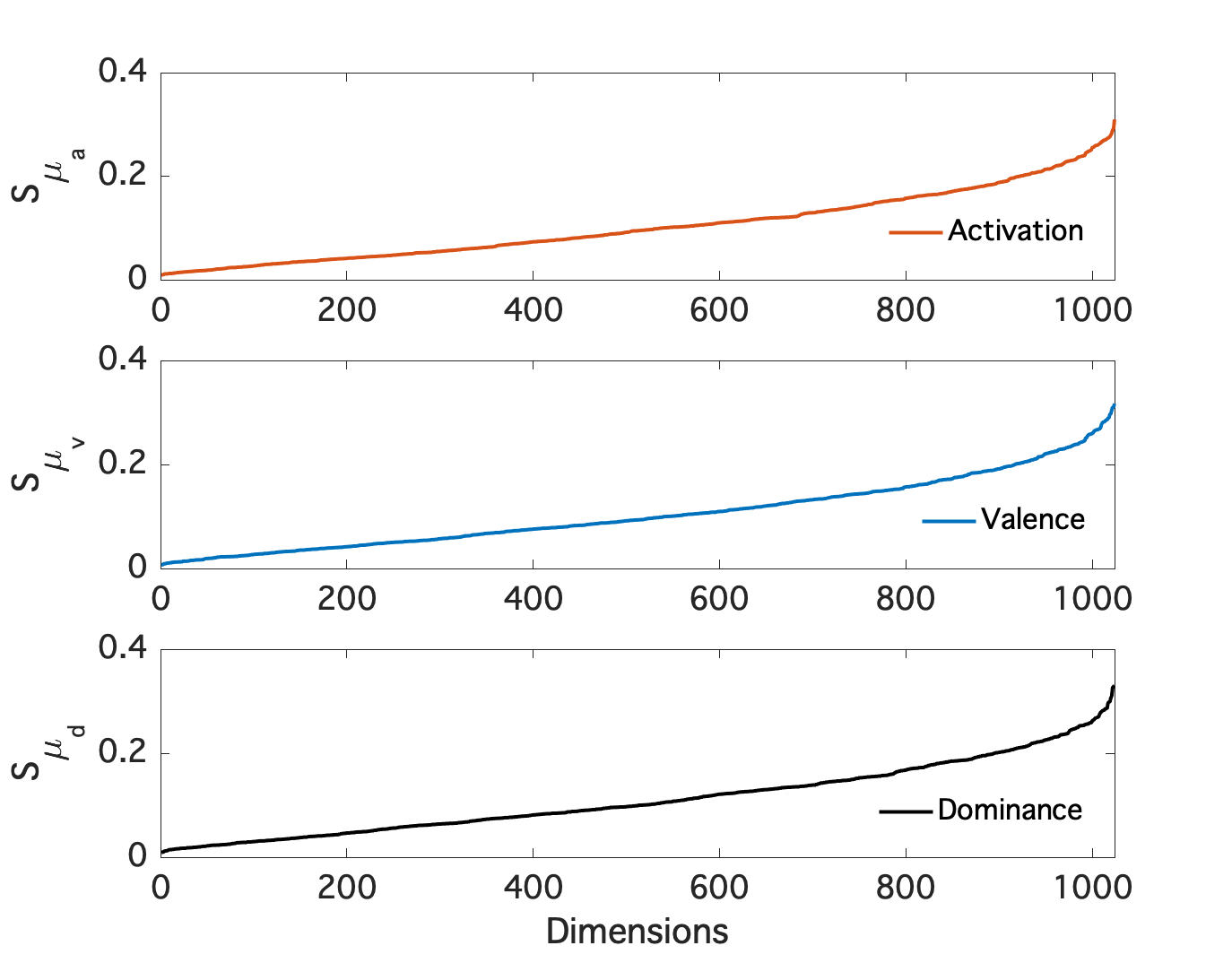}}
\vspace{-\baselineskip}
\end{minipage}
\caption{HUBERT Large saliency plot}
\label{fig:fig1}
\vspace{-\baselineskip}
\end{figure}

The pre-trained model embeddings have large dimensions. For example, both $HUBERT_L$ and $HUBERT_A$ have 1024 dimensions each, and when combined they result in a large input dimensional representation for the downstream emotion models. Prior studies \cite{srinivasan2021representation, mitra2023pre} have shown that when such large dimensional features are used to train speech emotion models, they result in small dimensional embedding space (in the range of $\approx$ 128 to 256). This finding indicates that dimensional emotion-relevant information in the input representations may reside in a subspace and the knowledge of that subspace can help to reduce the downstream model size. 
We investigate obtaining emotion-salient representations from the $HUBERT_L$ and $HUBERT_A$ representations, leveraging relationships between the input representation space and the targets. Prior work \cite{mitra2020investigation} has explored the input-output relationships of neurons to obtain neural saliency, and we use a similar idea to obtain the saliency of pre-trained model representations for dimensional emotion estimation. Let the $k^{th}$ representation of $N$ dimensional $HUBERT$ for an utterance $y$ be represented by a vector $H_{k,y} = [X_{1,k}, \dots, X_{i,k}, \dots, X_{M,k}]$, where $M$ denotes the sequence length. Let the mean affect labels for $y$ be $L_y \in \{\mu_{v,y}, \mu_{a,y}, \mu_{d,y}\}$, where ${\mu_{v,y}, \mu_{a,y}, \mu_{d,y}}$ represent the mean valence, activation and dominance scores for utterance $y$. $\overline{H}_{y,k}$ in eq. \ref{eq1} is obtained from $H_{y,k}$ by taking the mean across all the frames for utterance $y$. The cross-correlation based saliency ($CCS$) of $k^{th}$ dimension is given by 
\newcommand\normx[1]{\left\Vert#1\right\Vert}
\begin{equation}
S_{CCS} = \normx {\frac {Cov({\overline{H}_k},L)}{\sigma_{H_k}\sigma_L}} + \gamma_k
    \label{eq1}
\end{equation}

\noindent where eq. \ref{eq1} computes the absolute cross-correlation between labels $L \in \{\mu_{v}, \mu_{a}, \mu_{d}\} $ and embeddings ${\overline{H}_{k}}$ for dimension $k$ for all utterances in the \textit{training} set. $\gamma_k$ is the sum of the weighted cross-correlation between the $k^{th}$ dimension and all other dimensions, as shown in eq. \ref{eq2}
\begin{equation}
\gamma_k = \frac {1}{N-1} \sum_{j=1, j \ne k}^{N} w_j \normx {\frac {Cov({\overline{H}_k},{\overline{H}_j})}{\sigma_{{\overline{H}_k}}\sigma_{{\overline{H}_j}}}} \\ 
\label{eq2}
\end{equation}
\begin{center}
where, $w_j = \normx {\frac {Cov({\overline{H}_j},L)}{\sigma_{{\overline{H}_j}}\sigma_L}}$
\end{center}
\noindent  
\begin{equation}
\mu_{CCS} = \frac {1}{3}\sum_{L \in \{\mu_{v}, \mu_{a}, \mu_{d}\}} {S_{{CCS}_L}}.
    \label{eq3}
\end{equation}
In our experiments we have used $\mu_{CCS}$ given in eq. \ref{eq3} to select salient dimensions in pre-trained representations. We also investigated the use of mutual information instead of cross-correlation to obtain saliency information, where the cross-correlation terms in eq. \ref{eq1} and \ref{eq2} are replaced with mutual information, resulting in mutual information based saliency ($MIS$). In addition, we have explored principal component analysis ($PCA$) based dimensionality reduction of the input representation to compare with the proposed saliency based dimensionality reduction approaches. Figure \ref{fig:fig1} shows the saliency scores from $HUBERT_L$, across all three dimensional emotions, when sorted by saliency score (low to high from left to right). Note that around 20\% of the $HUBERT$ dimensions show salience $\leq 0.05$, which can be removed for dimensional emotion modeling. Figure \ref{fig:fig3} shows the scatter plot of the saliency scores for $HUBERT_A$ representations (N=1024 dim.), across valence, activation and dominance, where each circle represents a $HUBERT_A$ dimension, and the color-coded filled circles represent the top 60\% salient $HUBERT_A$ dimensions based on $CCS$.
\vspace{-\baselineskip}
\vspace{1mm}
\vspace{-\baselineskip}
\vspace{1mm}
\begin{figure}[h!]
\begin{minipage}[b]{1.0\linewidth}
  \centering
  \centerline{\includegraphics[width=7.3cm]{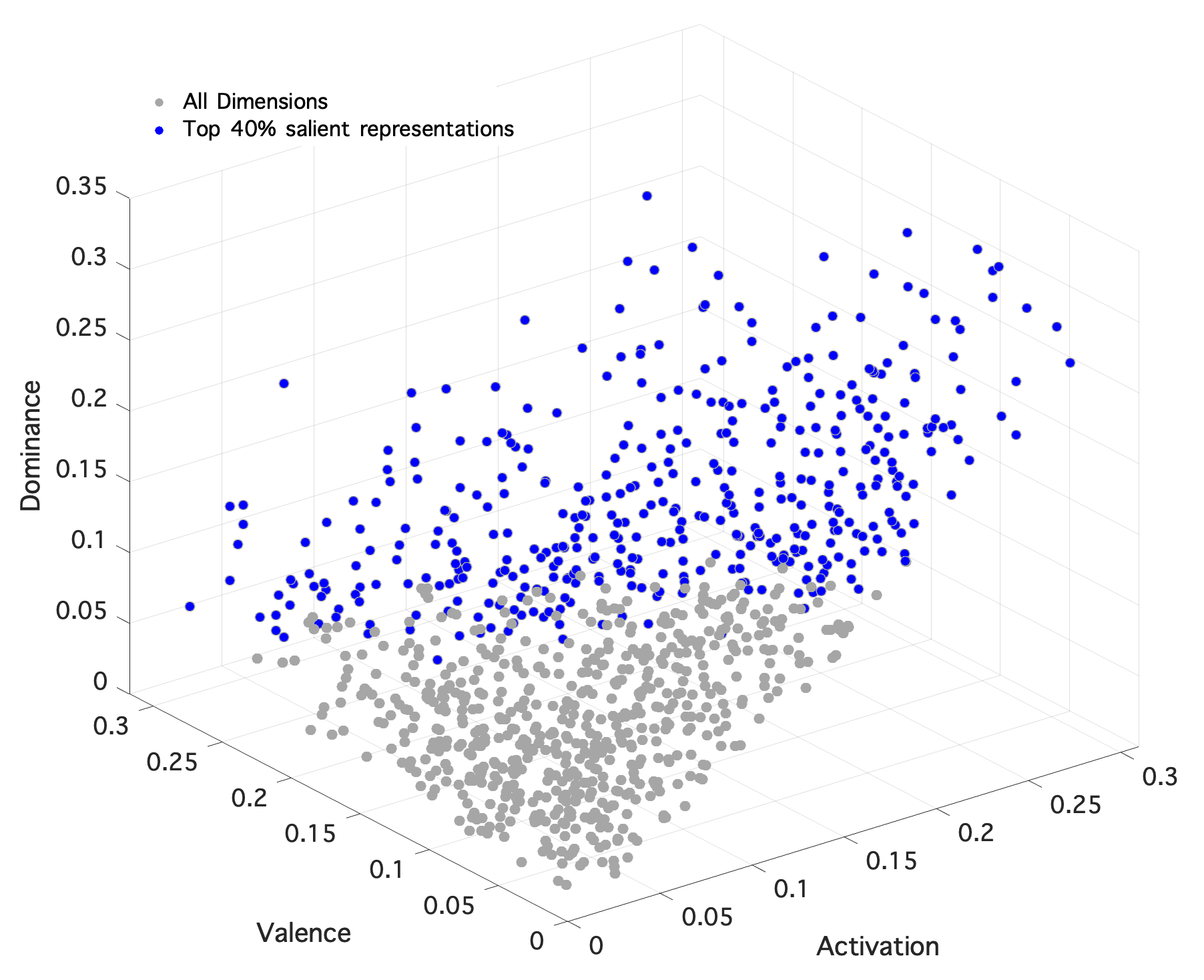}}
\vspace{-\baselineskip}
\end{minipage}

\caption{HUBERT ASR Saliency scatter plot across valence, activation and dominance dimensions}
\vspace{-2mm}
\label{fig:fig3}
\end{figure}
\vspace{-\baselineskip}
\vspace{2mm}
 
\subsection{Model Training}
\vspace{-2mm}
For dimensional emotion estimation modeling, we trained a network consisting of time-convolutional gated recurrent unit (TC-GRU) as shown in Fig. \ref{fig:fig4}. The model was trained with Train1.6 data (see Table \ref{tab:table1}, where the performance on a held-out Valid1.6 set was used for model selection and early stopping. Concordance correlation coefficient ($CCC$) \cite{lawrence1989concordance} is used as the loss function ($L_{ccc}$) (see eq. (\ref{eq4})), where ${L_{ccc}}$ is a combination ($\alpha=1/3$ and $\beta=1/3$) of $CCC$'s obtained from each of the three dimensional emotions. $CCC$ is defined by eq. (\ref{eq5}), where ${\mu _{x}}$ and ${\mu _{y}}$ are the means, ${\sigma _{x}^{2}}$ and ${\sigma _{y}^{2}}$ are the corresponding variances for the estimated and ground-truth variables, and ${\rho}$ is the correlation coefficient between those variables. The models are trained with a mini-batch size of 64 and a learning rate of 0.0005. The GRU network had two layers with 256 neurons each and an embedding size of 128, the convolution layer had a kernel size of 3.
\begin{multline}
  L_{ccc}= - (\alpha CCC_{v}+\beta CCC_{a}+(1-\alpha-\beta)CCC_{d}),
  \label{eq4}
\end{multline}
\begin{equation}
CCC = \frac {2\rho \sigma_x\sigma_y}{\sigma_x^2+\sigma_y^2 +(\mu_x-\mu_y)^2 }.
    \label{eq5}
\end{equation}

\begin{figure}[h]
\begin{minipage}[b]{1.0\linewidth}
  \centering
  \centerline{\includegraphics[width=3.2in]{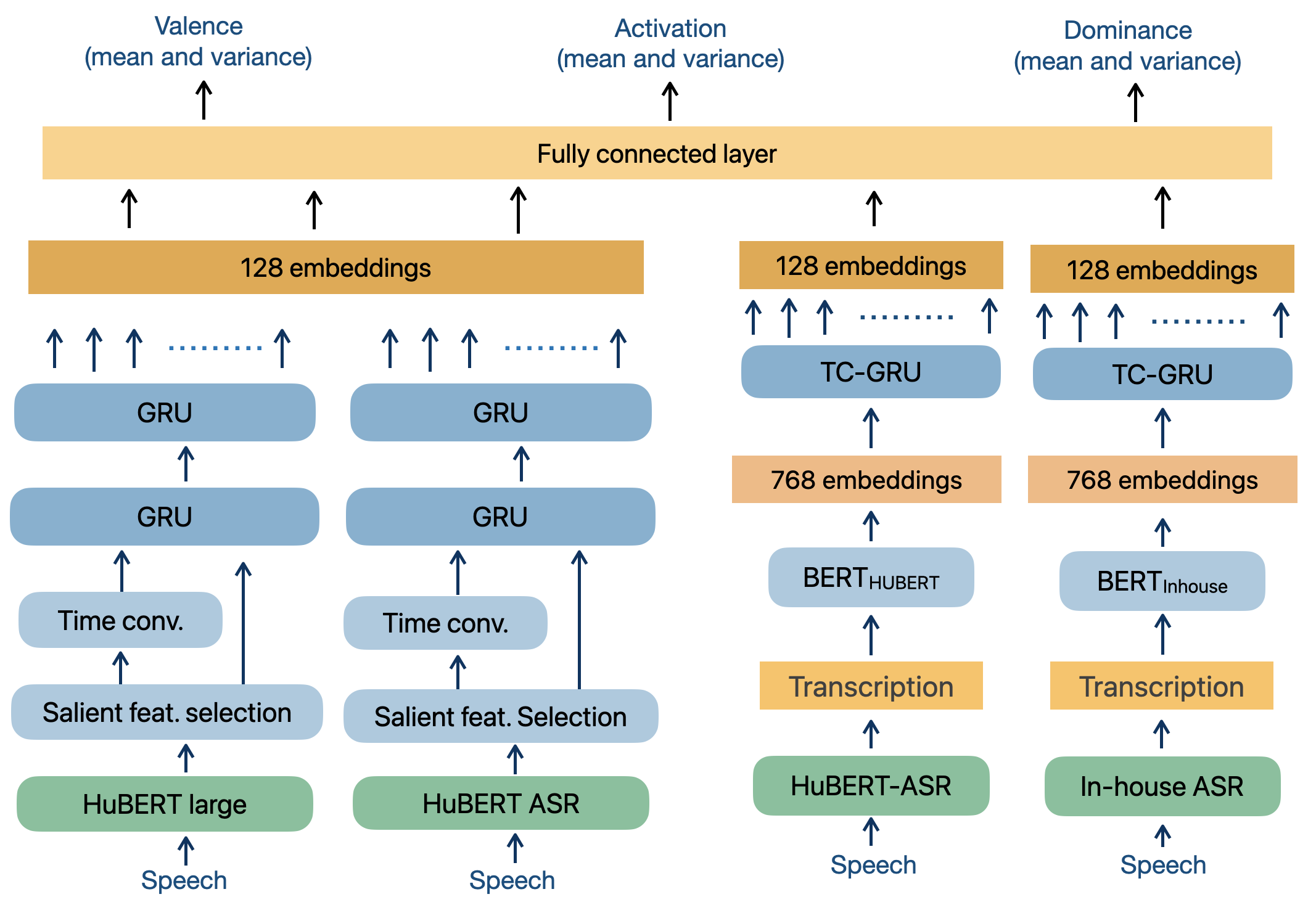}}
\end{minipage}
\vspace{-\baselineskip}
\caption{Multi-modal emotion estimation model}
\setlength{\belowcaptionskip}{0pt}
\label{fig:fig3}
\end{figure}
\vspace{-5mm}

\section{Results}
\vspace{-2mm}
\subsection{Emotion estimation}
\vspace{-2mm}
We trained a multi-modal emotion estimation model (see Fig \ref{fig:fig3} with the following inputs: embeddings from (i) $HuBERT_L$ and (ii) $HuBERT_A$, and lexical embeddings from (iii) BERT using transcriptions generated by $HuBERT_{A}$ and (iv) BERT embeddings from an in-house ASR generated transcriptions. The performance of the dimensional emotion estimation model is shown in Table \ref{tab:table2}. 

\begin{table}[h]
\caption{Dimensional emotion estimation Concordance Correlation Coefficient ($CCC$) performance of models: (1) without target variance (\texttt{w/o}) and with target variance (\texttt{w}), (ii) with removal of less salient (using $CCS$) input representations, and (iii) under noisy conditions}
\vspace{2mm}

\centering
    \begin{adjustbox}{width=0.5\textwidth,center}
        \begin{tabular}{@{}ll@{\hspace{0.5pt}}ccccc@{}}
            \toprule
                  & \multicolumn{1}{c}{target} &\multicolumn{1}{c}{\% Input} & \multicolumn{1}{c}{$\downarrow$ \% Rel.} & & \\
              & {variance} & {Reps.} & {model size} & {Act} & {Val} & {Dom} \\
              
            \toprule
            \multirow{5}{*}{$\mathbf{Eval1.3}$} & \multicolumn{1}{l}{\texttt{w/o}} & \multicolumn{1}{l}{$100\%$} & - & \multicolumn{1}{c}{0.77}                                    &\multicolumn{1}{c}{0.68} & \multicolumn{1}{c}{0.69} \\\cline{2-7}
            
                                              & \multicolumn{1}{l}{\texttt{w}} & \multicolumn{1}{l}{$100\%$} & - & \multicolumn{1}{c}{0.78} & \multicolumn{1}{c}{0.69} & \multicolumn{1}{c}{0.70}\\\cline{2-7}
                                             & \multicolumn{1}{l}{\texttt{w}} & \multicolumn{1}{l}{$80\%$} & \multicolumn{1}{c}{$16\%$} & \multicolumn{1}{c}{0.77} & \multicolumn{1}{c}{0.68} & \multicolumn{1}{c}{0.69}\\\cline{2-7}
                                             & \multicolumn{1}{l}{\texttt{w}} & \multicolumn{1}{l}{$60\%$} & \multicolumn{1}{c}{$32\%$} & \multicolumn{1}{c}{0.76} & \multicolumn{1}{c}{0.67} & \multicolumn{1}{c}{0.68}\\\cline{2-7}
                                             & \multicolumn{1}{l}{\texttt{w}} & \multicolumn{1}{l}{$40\%$} & \multicolumn{1}{c}{$49\%$} & \multicolumn{1}{c}{0.75} & \multicolumn{1}{c}{0.66} & \multicolumn{1}{c}{0.67} \\
                                             \bottomrule
            \multirow{5}{*}{$\mathbf{Eval1.6}$} & \multicolumn{1}{l}{\texttt{w/o}} & \multicolumn{1}{l}{$100\%$} & - & \multicolumn{1}{c}{0.74}                                    &\multicolumn{1}{c}{0.66} & \multicolumn{1}{c}{0.66} \\\cline{2-7}
            
                                              & \multicolumn{1}{l}{\texttt{w}} & \multicolumn{1}{l}{$100\%$} & - & \multicolumn{1}{c}{0.75} & \multicolumn{1}{c}{0.67} & \multicolumn{1}{c}{0.67}\\\cline{2-7}
                                             & \multicolumn{1}{l}{\texttt{w}} & \multicolumn{1}{l}{$80\%$} & \multicolumn{1}{c}{$16\%$} & \multicolumn{1}{c}{0.75} & \multicolumn{1}{c}{0.66} & \multicolumn{1}{c}{0.67}\\\cline{2-7}
                                             & \multicolumn{1}{l}{\texttt{w}} & \multicolumn{1}{l}{$60\%$} & \multicolumn{1}{c}{$32\%$} & \multicolumn{1}{c}{0.74} & \multicolumn{1}{c}{0.65} & \multicolumn{1}{c}{0.66}\\\cline{2-7}
                                             & \multicolumn{1}{l}{\texttt{w}} & \multicolumn{1}{l}{$40\%$} & \multicolumn{1}{c}{$49\%$} & \multicolumn{1}{c}{0.73} & \multicolumn{1}{c}{0.64} & \multicolumn{1}{c}{0.65} \\\bottomrule
            \multirow{4}{*}{$\mathbf{Eval1.6_{25dB}}$} & \multicolumn{1}{l}{\texttt{w}} & \multicolumn{1}{l}{$100\%$} & - & \multicolumn{1}{c}                                    {0.72} & \multicolumn{1}{c}{0.66} & \multicolumn{1}{c}{0.63}\\\cline{2-7}
                                             & \multicolumn{1}{l}{\texttt{w}} & \multicolumn{1}{l}{$80\%$} & \multicolumn{1}{c}{$16\%$} & \multicolumn{1}{c}{0.72} & \multicolumn{1}{c}{0.65} & \multicolumn{1}{c}{0.63}\\\cline{2-7}
                                             & \multicolumn{1}{l}{\texttt{w}} & \multicolumn{1}{l}{$60\%$} & \multicolumn{1}{c}{$32\%$} & \multicolumn{1}{c}{0.72} & \multicolumn{1}{c}{0.65} & \multicolumn{1}{c}{0.62}\\\cline{2-7}
                                             & \multicolumn{1}{l}{\texttt{w}} & \multicolumn{1}{l}{$40\%$} & \multicolumn{1}{c}{$49\%$} & \multicolumn{1}{c}{0.70} & \multicolumn{1}{c}{0.64} & \multicolumn{1}{c}{0.60} \\
                                             \bottomrule
            \multirow{4}{*}{$\mathbf{Eval1.6_{15dB}}$} & \multicolumn{1}{l}{\texttt{w}} & \multicolumn{1}{l}{$100\%$} & - & \multicolumn{1}{c}                                   {0.69} & \multicolumn{1}{c}{0.64} & \multicolumn{1}{c}{0.61}\\\cline{2-7}
                                             & \multicolumn{1}{l}{\texttt{w}} & \multicolumn{1}{l}{$80\%$} & \multicolumn{1}{c}{$16\%$} & \multicolumn{1}{c}{0.69} & \multicolumn{1}{c}{0.64} & \multicolumn{1}{c}{0.60}\\\cline{2-7}
                                             & \multicolumn{1}{l}{\texttt{w}} & \multicolumn{1}{l}{$60\%$} & \multicolumn{1}{c}{$32\%$} & \multicolumn{1}{c}{0.68} & \multicolumn{1}{c}{0.64} & \multicolumn{1}{c}{0.60}\\\cline{2-7}
                                             & \multicolumn{1}{l}{\texttt{w}} & \multicolumn{1}{l}{$40\%$} & \multicolumn{1}{c}{$49\%$} & \multicolumn{1}{c}{0.66} & \multicolumn{1}{c}{0.63} & \multicolumn{1}{c}{0.56} \\\bottomrule
            \multirow{4}{*}{$\mathbf{Eval1.6_{5dB}}$} & \multicolumn{1}{l}{\texttt{w}} & \multicolumn{1}{l}{$100\%$} & - & \multicolumn{1}{c}                                   {0.54} & \multicolumn{1}{c}{0.57} & \multicolumn{1}{c}{0.45}\\\cline{2-7}
                                             & \multicolumn{1}{l}{\texttt{w}} & \multicolumn{1}{l}{$80\%$} & \multicolumn{1}{c}{$16\%$} & \multicolumn{1}{c}{0.54} & \multicolumn{1}{c}{0.57} & \multicolumn{1}{c}{0.44}\\\cline{2-7}
                                             & \multicolumn{1}{l}{\texttt{w}} & \multicolumn{1}{l}{$60\%$} & \multicolumn{1}{c}{$32\%$} & \multicolumn{1}{c}{0.51} & \multicolumn{1}{c}{0.59} & \multicolumn{1}{c}{0.43}\\\cline{2-7}
                                             & \multicolumn{1}{l}{\texttt{w}} & \multicolumn{1}{l}{$40\%$} & \multicolumn{1}{c}{$49\%$} & \multicolumn{1}{c}{0.48} & \multicolumn{1}{c}{0.59} & \multicolumn{1}{c}{0.39} \\\bottomrule

        \end{tabular}
    \end{adjustbox}
    \vspace{ -5 mm}
    \label{tab:table2}
\end{table}

\begin{table}[h!]
\caption{Dimensional emotion estimation $CCC$ after selecting 80\% of the input representations using $CCS$, $MIS$ and $PCA$ based approaches}
\begin{adjustbox}{width=0.4\textwidth,center}
\centering
   \begin{tabular}{lcccccc}
     \toprule
     \multirow{2}{*}{\textbf{System}} &
       \multicolumn{3}{c}{\textbf{Eval1.3}} &
       \multicolumn{3}{c}{\textbf{Eval1.6}} \\
       & {act} & {val} & {dom} & {act} & {val} & {dom} \\
       \midrule
     $ PCA$ & 0.75 & 0.67 & 0.67 & 0.73 & 0.65 & 0.65  \\
     $ MIS$ & 0.77 & 0.68 & 0.69 & 0.74 & 0.66 & 0.66 \\
     $ CCS$ & 0.77 & 0.68 & 0.69 & 0.75 & 0.66 & 0.67 \\
     \bottomrule
   \end{tabular}
   \label{tab:table3}
   \end{adjustbox}
   \vspace{ -5 mm}
\end{table}

We investigated the model performance when: (i) incorporating target variance for both Eval1.3 and Eval1.6 test sets across all dimensions; (ii) reducing the model size as a consequence of input representation selection based on saliency ($CIS$); and (iii) under noise conditions as shown in Table~\ref{tab:table2}. Incorporating target variance helped to improve emotion estimation performance compared to the model not using it, where 1 to 2 \% relative improvement (statistically significant) in $CCC$ was observed across all dimensional emotions for both Eval1.3 and Eval1.6 test-sets. Removing 20\%, 40\%, and 60\% of input representations based on saliency (\% Input Resps. = 80\%, 60\%, and 40\%) translated to a relative 16\%, 32\%, and 49 \% reduction in model size, where the baseline model size was 15.6MB. This reductions in model sizes corresponded to relative decreases in $CCC$ from 1 to 6\% (statistically significant). Note that the impact of noise on the performance of models based on salient representations was minimal. Table~\ref{tab:table2} shows that for 16\% reduction the model's performance is comparable, while for 32\% and 49\% reduction, there is some regression, compared to the baseline.

\begin{figure}[h!]
\begin{minipage}[b]{1.0\linewidth}
  \centering
  \centerline{\includegraphics[width=6.9cm]{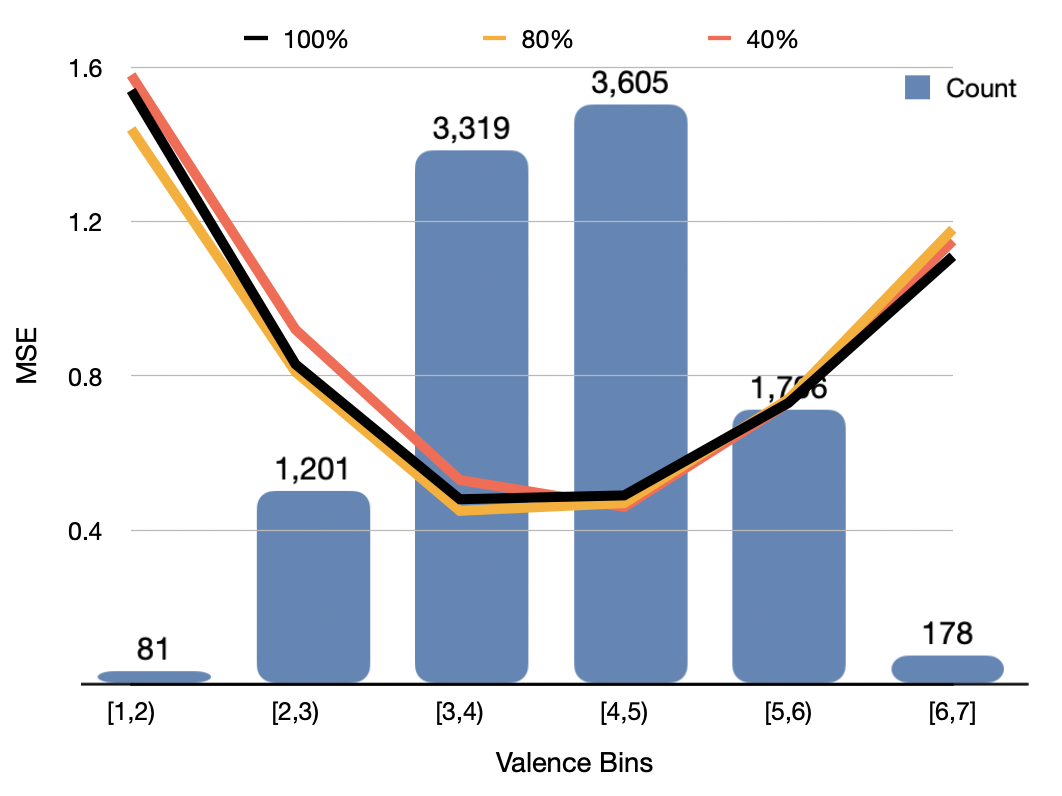}}
\end{minipage}
\vspace{-5mm}
\caption{Mean squared error (MSE) for valence estimation at different target bins from models trained with full representation (100\%), saliency-based feature selection (80\% and 40\%)}
\label{fig:fig4}
\vspace{-\baselineskip}

\end{figure}

When comparing the performance of models trained with full feature representations against those employing saliency-based feature selection (\% Input Resps. = 80\%, 60\%, and 40\%), it is observed that the mean squared error (MSE) for estimating three dimensions across various 7-point Likert score bins remains consistent despite the reduction in model size. A selected illustrative example of this trend with valence estimation is shown in Figure~\ref{fig:fig4}. Note that MSE increased for bins with less data points.

Finally, we compared emotion estimation performance of models trained with 80\% of pre-trained model dimensions, where the dimensionality reduction was performed with saliency-based ($CCS$ and $MIS$) and $PCA$ based approaches. Table \ref{tab:table3} shows that both the saliency based approach ($CCS$ and $MIS$) performed better than $PCA$. Saliency based approaches reduced dimension by removing less-salient input feature dimensions, while retaining the temporal structure of the residual features. On the other hand, $PCA$ uses a singular value decomposition of the data to project it to a lower dimensional space and in that process does not retain the temporal structure of each of the feature dimensions. The lack of temporal structure in $PCA$ based dimensionality reduction attributes to its lower performance (see table \ref{tab:table3}) when compared to the saliency based approaches. Note that both $CCS$ and $MIS$ based dimensionality reduction approaches performed similarly (see Table \ref{tab:table3}), where $CCS$ performed better than $MIS$ only for Eval1.6 activation and dominance estimation.

\vspace{-2mm}
\section{Conclusions}
In this work, we investigated if it is possible to select salient representations from a pre-trained model for the task of dimensional emotion estimation from speech. We observed that such selection can result in substantial model size reduction, without significant regression in performance. We evaluated the model performance under unseen acoustic distortions, and observed that saliency-driven representation selection helped the reduced-sized models to perform as-good-as the large baseline models, without substantial regression in model's generalization capacity. In the future, we plan to investigate saliency-driven pruning of the downstream models, to investigate if the model size can be reduced even further without regression in performance and generalization capacity. 

\bibliographystyle{IEEEbib}
\bibliography{strings,refs}

\end{document}